\documentclass[12pt]{iopart}
\usepackage{epsfig}
\begin{document}
\title{Non locality and causal evolution in QFT}
\author{F Buscemi \dag\ and G  Compagno \ddag}
\address{\dag\ CNR-INFM National S3 Research Center, via Campi 213/A, I-Modena 41100, Italy and
and Dipartimento di Fisica, Universit\`{a} di Modena e Reggio
Emilia, Modena, Italy}
\address {\ddag\ CNR-INFM, and
Dipartimento di Scienze Fisiche ed Astronomiche
dell'Universit\`{a} di Palermo, Via Archirafi 36, 90123 Palermo,
Italy} \eads{\dag\ \mailto{buscemi.fabrizio@unimore.it} \ddag\
\mailto{compagno@fisica.unipa.it}}

\begin{abstract}

Non locality appearing in QFT during the free evolution of localized field states and in the Feynman propagator function is analyzed. It is shown to be connected to the initial non local properties present at the level of quantum states and then it does not imply a violation of Einstein's causality. Then it is investigated a simple QFT system with interaction, 
consisting of a classical source coupled linearly to a quantum scalar field, that is exactly solved. The expression for the time evolution 
of the state describing the system is given. The expectation value of any arbitrary ``good''  local observable, expressed as a function of 
the field operator and its space and time derivatives, is obtained explicitly at all order in the field-matter coupling constant. 
These expectation values have a source dependent part that is shown to be always causally retarded, 
while the non local contributions are source independent and related to the non local properties of zero point vacuum fluctuations.

\end{abstract}

\submitto{\JPB} 

\section{Introduction}

In non relativistic  Quantum Electrodynamics (QED) non  locality has initially been studied  in the context of the energy transfer between couple of atoms, the so-called Fermi
problem \cite{Ferm}, and then in other systems  \cite{kar,Pri2,Mate,Pow} giving rise to controversial interpretations \cite{Rubin,Ber,Ber2,Bus,Buc,milo,Com3}. 
In relativistic quantum mechanics Hegerfeldt's theorem (HT)\cite{Heg3} associates the appearance of non locality to the condition of positivity of  the energy. 
In fact wavefunctions initially localized in finite space regions develop with time non zero contributions 
and also finite expectation values for some localization operators  \cite{Heg,Hege} at spacelike distances from the localization region, apparently at variance with Einstein's causality.

The use of localization or of position operators to determine if the appearance of non locality gives rise to observable effects appears, however, in relativistic quantum mechanics beset with difficulties\cite{Pere,Rose}.
 In Quantum Field Theory (QFT), where non local quantities as the Feynman propagator already show up for free fields \cite{Rubin},
 the standard QFT techniques for interacting fields relate causality to the analytic and unitary properties of the S-matrix \cite{Sto}. 
This, with the use of perturbation theory, makes the analysis of causality a ``difficult thing in QFT''  \cite{Vel}.  
In fact the most  natural approach to analyze the consequences of HT would be to follow the time development of the systems  as is usually done, within perturbation theory,  in non relativistic QED.  
There however other kind of difficulties appears  in the description  of  the sources or the detectors of the field that propagates. 
These are atoms that to be  approximated  as  effectively point like  must be initially described by bare states (that allow structure localized in small space regions)
 if  instead described by dressed states  they are inevitably complex structures of atoms plus field extended in space \cite{Pri2,Pri1,libro}. 
Another limitation is that here again  the analysis can  be limited to the lowest orders of perturbation theory.

To overcome some of these difficulties, explicit simplified interacting QFT models with classical sources have been used  \cite{kar,Mate,Bus}.These 
 permit to follow in detail the time evolution, are not beset with the problems in the definition of the initial states used
 to describe the sources and cab be solved non  perturbatively \cite{Mate}. In one of these models, consisting of the electromagnetic field interacting with a classical source, 
 it has been shown that while the expectation value of the field energy density behaves causally  non local effects appear at the level of quantum states, giving ``acausal'' 
 effects that are not however ``observable''  at a classical level \cite{kar}. 
In another model, of a scalar field interacting with a classical instantaneous and point like source  and within second order perturbation theory, 
it has been shown that the evolution of one- and two-point field operators are non local. Non locality however appears either in source independent terms,  
and may be attributed to zero point vacuum fluctuations, or in source dependent terms and may then be attributed to the use of operators 
whose commutator is not zero for space like distances and that, not satisfying the  microcausality principle,  cannot be considered ``“good'' operators \cite{Bus}.

Because, for a given model,  a proof of causality would be required to held at any order and for any ``good'' local observable, 
the aim of this paper is to extend the previous perturbative results \cite{Bus}  to all orders and to explicitly calculate the form of the expectation value 
of any ``good'' operator, expressed locally in terms of the fields,  generalizing the results of references \cite{kar} and \cite{Mate}.
 Moreover we want also that our analysis of the causality be independent of a specific assigned evolution of the source. 
For this purpose in the following we shall consider the QFT model of a quantum scalar field linearly interacting with a classical scalar source 
arbitrarily extended in space and varying in time although localized in an finite space-time region. We aim also to show that 
the appearance of non locality already at the level of the free field in QFT theory may be attributabed  to the fact that, to create localized states, 
the operators are used that do not create effectively  point like states. The paper is structured as follows: in Sec.\ref{intro}  the appearance of non locality for a free quantum scalar field is analyzed. 
In Sec.~\ref{intro2} and \ref{intro3} the time evolution of the state of the scalar field coupled to a source and the expectation value on it of arbitrary local ``good''  operators is obtained. 
Finally in Sec.~\ref{intro4} we shall comment the results obtained.

\section{Single particle amplitude evolution} \label{intro}
Locality for quantum fields interacting with source has been investigated by analyzing the expectation
values of relativistic localization operators \cite{kar,Pri2,Pri1}.
Here we want to contend  that  non locality that appears  in these treatments may be held to have the same origin of the non local characteristic shown by the standard
two-point  functions and  by the single particle
amplitude that  already shows  up in the free field case. In particular we shall  analyze the case of quantum
scalar field $\Phi(x)$, $x \equiv (\textbf{x},x^{0}=t)$, that expressed  in terms of its positive
and negative frequency part is:
\begin{equation} \label{modena}
\Phi(x)=\Phi_{+}(x)+\Phi_{-}(x)
\end{equation}
where, taking $\hbar=1$ and $c=1$,
\begin{equation} \label{moden2}
\Phi_{+}(x)=\frac{1}{(2\pi)^{3/2}}\int
\frac{\rmd^{3}\textbf{k}}{2\omega}\,a(\textbf{k}) e^{ -ik\cdot
x}\quad\textrm{and}\quad \Phi_{-}(x)=\Phi_{+}^{\dag}(x),
\end{equation}
with  $\omega= \sqrt{|\textbf{k}|^{2}+m^{2}}$  and   $a(\textbf{k})$,
$a^{\dag}(\textbf{k})$ being  respectively the usual annihilation and
creation operators that satisfy  the relativistic commutator rules:
\begin{equation}
\left[ a(\textbf{k}),a^{\dag}(\textbf{k}')\right]=2\omega
\delta^{3}(\textbf{k}-\textbf{k}').
\end{equation}
The state  $\Phi(x)|0 \rangle$ generated by the action of the
field operator on the vacuum field state $|0\rangle$,  using
equations (\ref{modena}) and  (\ref{moden2}), is
\begin{equation}\label{vend}
\Phi(x)|0 \rangle=\Phi_{-}(x)|0 \rangle=\frac{1}{(2\pi)^{3/2}}\int
\frac{\rmd^{3}\textbf{k}}{2\omega}\,e^{ ik\cdot
x}\,a^{\dag}(\textbf{k}) |0\rangle
\end{equation}
where  $\Phi_{+}(x) |0\rangle=0$ has been taken into account.
 $\Phi(x)|0 \rangle$ is a linear superposition of single
 particle states  $a^{\dag}(\textbf{k}) |0\rangle=|\textbf{k}\rangle$,
  eigenstates   of the momentum operator with eigenvalue
 $\textbf{k}$. Except for the factor $1/2\omega$,
  expression (\ref{vend}) corresponds to the non relativistic expansion in terms of momenta of the eingenstate position
 $|\textbf{x}\rangle$.  Thus one is lead to assume \cite{Pes}  that $\Phi(x)|0
 \rangle$ represents the   ``localized'' state where a single particle is
 created at position $\textbf{x}$. Apparently a confirmation is given by the fact that  projecting  this
 state on momentum eigenstate $|\textbf{k}\rangle$, we obtain
 \begin{equation}
 \langle 0|\Phi(x)|\textbf{k}\rangle=\frac{1}{(2\pi)^{3/2}}\int
\frac{\rmd^{3}\textbf{k}'}{2\omega}\, e^{ -ik'\cdot x}\, \langle
0|a(\textbf{k}')a^{\dag}(\textbf{k}) |0\rangle= \frac{e^{ -ik\cdot
x}}{(2\pi)^{3/2}}
\end{equation}
and thus  interpret  the above expression as the space
representation of the single particle wavefunction of the state
$|\textbf{k}\rangle$, just as in non relativistic  quantum
mechanics  $\langle \textbf{x}|\textbf{k}\rangle \propto e^{
-ik\cdot x}$ is the wavefunction of  $|\textbf{k}\rangle$. However
as we shall see  this interpretation cannot be held too strictly.
In fact the  two-point function given by the scalar product
between $\Phi(x)|0 \rangle$ and $\Phi(y)|0 \rangle$ is:
\begin{equation}\label{berluska}
\langle 0|\Phi(x)\Phi(y)| 0 \rangle=\frac{1}{(2\pi)^{3}}\int
\frac{\rmd^{3}\textbf{k}}{2\omega}\, e^{ -ik\cdot (x-y)}=
i\Delta_{+}(x-y)
\end{equation}
where  $\Delta_{+}(x-y)$ is the positive frequency propagator part
of $\Delta$ function. Equation~(\ref{berluska}) is  interpreted as
the probability amplitude  of finding at spacetime point  $y$ a
particle created at point  $x$. The explicit form of
$\Delta_{+}(x-y)$  \cite{Pes,Bjo2} shows that it is not zero for
$x-y$ spacelike and does not tend to $3$-dimension  $\delta$
function when $x_{0}\rightarrow y_{0}$. In fact:
\begin{equation}\label{jac}
\lim_{x_{0}\rightarrow y_{0},|x-y|\rightarrow \infty} \Delta_{+}(x-y)\propto
\left \{
\begin{array}{ccc}
e^{ -m|x-y|} &\quad \textrm{for}\quad m \neq
0 \\
 \frac{1}{|x-y|^{2}} & \quad \textrm{for}\quad
m = 0
\end {array} \right . .
\end{equation}

Considering instead the  equation~(\ref{berluska}) as the scalar
product of the  the states $\Phi(x)|0 \rangle$ and $\Phi(y)|0
\rangle$,  describing a single particle localized respectively  at
spacetime points $x$ and $y$,   we see  a spatial overlap for
spacelike distances. This peculiar behaviour can be avoided  if we
give up the interpretation of $\Phi(x)|0 \rangle$ as a state where
a particle is created in $\textbf{x} $ and is there $\delta$ like
localized, but instead that is represents  a state, extending over
the whole space,  centered at $\textbf{x}$. This interpretation is
also supported by evaluating on this state ``good'' local
observables that satisfy the microcausality principle. In
particular taking the  scalar field energy density at $x$:
\begin{equation} \label{dink}
\!\mathcal{H}(x)\!=\!\frac{1}{2}\!\left(\!|\nabla
\Phi(x)|^{2}\!\!+\!\dot{\Phi}(x)^{2}\!+\!m^{2}\Phi^{2}(x)\!\right)
\end{equation}
it has the   expectation value on the ``localized''  state at $y$
$\Phi(y)|0\rangle$
\begin{eqnarray}\fl  \langle 0|\Phi(y)\mathcal{H}(x)\Phi(y)| 0 \rangle\!= -2m^{2}\Delta_{+}(x-y)\Delta_{+}(y-x)\nonumber \\ \fl   -2\partial_{x_{0}}\Big(\Delta_{+}(x-y)\Delta_{+}(y-x)\Big) -2\nabla_{x}\Big(\Delta_{+}(x-y)\Delta_{+}(y-x)\Big){}  \nonumber \\
 \fl    - \lim_{s\rightarrow 0}\Big(m^{2}\Delta_{+}^{2}(s)+\Delta_{+}(s)\,\partial_{s_{0}}^{2}\Delta_{+}(s)+\Delta_{+}(s)\,\nabla_{s}^{2}\Delta_{+}(s)\Big)
\end{eqnarray}
which, from the properties of the $\Delta_{+}$ two-point function
given by the expression~(\ref{jac}),  can be seen  to have
contributions different from zero over the whole space and not a
$\delta$ behaviour.

This interpretation of the    state $\Phi(x)|0 \rangle$
 permits
us to explain  the appearance of non local effects present  in the Feynman propagator $\Delta_{F}$:
\begin{eqnarray}\label{xxx}
\fl \Delta_{F}(x-y)&=&\langle 0|  T \{\Phi(x)\Phi(y)\}| 0 \rangle
\nonumber \\
& = &\Theta(x_{0}-y_{0})\langle 0|  \Phi(x)\Phi(y)| 0
\rangle+\Theta(y_{0}-x_{0})\langle 0|  \Phi(y)\Phi(x)| 0 \rangle,
\end{eqnarray}
 $T$ being the time ordering operator, with each of the terms appearing
 in the equation~(\ref{xxx}) is  usually  considered as the amplitude probability of
propagation  of particles localized from  $y$ to $x$  (or vice
versa). Following our previous discussion one can  interpret it as
the amplitude of going from the non localized  state centered at y
to the corresponding centered at $x$. It may therefore be expected,
because these  states are extended in space, that the overlap among these  may be non zero also for $|x-y|$ spacelike.

It may also be of interest to verify if the one particle states $\Phi(x)|0 \rangle$, on which
the  local  expectation value
of the field energy density is non local, may instead  satisfy the relativistic notion of localization adopted
 in relativistic quantum mechanics. In this framework the second quantized position
operator $\textbf{x}= i\nabla_{\textrm{\scriptsize{k}}}$ is not
anymore hermitian and this has lead  to  introduce  new definition
for the position operator and  the   states localized at a given
position \cite{kar,Pri2,Pri1}. In particular the Newton-Wigner
(NW) localized states   are single particle states    obtained
when the ``localization'' creation operator $a_{NW}^{\dag}(x)$
acts on vacuum state $| 0 \rangle $ \cite{Newi,Fle}. This operator
is related   to  the negative frequency part of the field operator
$\Phi(x)$ by  a non local integral transformation  \cite{Fle}:

\begin{equation}\label{opnw1}
a_{NW}^{\dag}(x)=\int
d^{3}\textbf{x}'K_{NW}(|\textbf{x}-\textbf{x}'|)\Phi_{-}(\textbf{x}',x_{0})
\end{equation}
where the  Kernel $K_{NW}$ defined as
\begin{equation}
K_{NW}(mR)=\frac{1}{(2\pi)^{3}}\int
d^{3}\textbf{k}\sqrt{2\omega}\,e^{i\textbf{\scriptsize{k}}\cdot\textbf{\scriptsize{R}}}
\end{equation}
has this asymptotic behaviour when $R$ tends to infinity
\begin{equation} \lim
_{R\rightarrow\infty}{K_{NW}}(R)\rightarrow \left \{
\begin{array}{ccc}
 cost
\frac{m^{7/2}}{(mR)^{9/4}}e^{-mR},&\quad \textrm{for}\quad m \neq
0 \\
 & \\
 cost R^{-7/2},&\quad \textrm{for}\quad m \neq
0
\end{array}\right ..
\end{equation}
 Thus   $K_{NW}$  extends,   for massive fields  ($m\neq 0$),  in a
space region of dimension comparable to Compton wavelength $1/m$,
while   for massless fields  $m=0$  it decreases  for large
distance with the  power law.
 In this context  the NW state localized at  $\textbf{x}$ at time $x_{0}$
is  written as
\begin{equation}
|X _{NW}\rangle= a_{NW}^{\dag}(x)|0\rangle = \frac{1}{(2 \pi)^{3/2}} \int \frac{d^{3}\textbf{k}}{\sqrt{2\omega}}\,\,a^{\dag}(\textbf{k})|0\rangle
\end{equation}
and its  scalar product with a  NW state localized at
$\textbf{y}$  at   $x_{0}=y_{0}$ may be shown to be:
\begin{equation}\label{neg}
\langle X_{NW}|Y _{NW}\rangle= \delta ^{3}(\textbf{x}-\textbf{y})
\quad \textrm{for}
\end{equation}
 therefore satisfying  the standard non relativistic condition for a $\delta$ localized state.
To examine   if the state, created from vacuum by the field operator,  $\Phi(x)|0 \rangle$
may be considered localized in NW sense,  we take its scalar product with state   $|Y _{NW}\rangle$  obtaining:
\begin{eqnarray} \label{anj}
\langle 0 |\Phi(x)|Y _{NW}\rangle &=& i \int
d^{3}\textbf{x}'K_{NW}(m|\textbf{y}-\textbf{x}'|)\Delta_{+}(x-x') {} \nonumber \\
&= & \frac{1}{(2\pi)^{3}}\int \frac{d^{3}\textbf{k}}{\sqrt{2\omega}}
e^{-ik\cdot(x-y)}=\psi _{y}^{NW}(x)
\end{eqnarray}
where   interestingly $\psi_{y}^{NW}(x)$ is just  the NW expression for
first quantized relativistic state localized  at $\textbf{y}$, which   differs from zero  for
 $ (x-y)$ spacelike and   does not reduce to $3$-dimension $\delta$ function even for $x_{0}=y_{0}$.
 Thus the state $\Phi(x)|0 \rangle$  cannot be even considered
localized in  the NW sense. This is another indication  that one
particle states created from vacuum by a direct application of the
field operator cannot represent physical situations with   the
single particle  restricted to an arbitrarily small region of
space.

As  known  the use of NW  localized states leads with time to the
appearance of non  local terms. In fact  the scalar product of the
NW state localized at $\textbf{y}$ at  $ x_{0}> y_{0}$ when
projected on the NW state localized at $\textbf{x}$  at   $ x_{0}$
is
\begin{equation}\label{vicinio}
\langle X_{NW}|Y _{NW}\rangle
=-2\partial_{x_{0}}\Delta_{+}(x-y)\neq \delta
^{3}(\textbf{x}-\textbf{y})
\end{equation}
 differing  from zero for spacelike distances. So a single particle state initially NW localized spreads
 non causally with time and  this result can be interpreted as an example of the HT.

\section{The  QFT model with source}\label{intro2}
In the previous section we have seen that in  QFT non locality
shows up at the level of single particle states and of transition
amplitude because the action of the free field operator $\Phi(x)$
on the vacuum effectively generates non local states. It seems
relevant to inquire if this non local behaviour may be observed
and this leads to consider situations where the the field
quantities are generated and detected. To this purpose we shall
explicitly examine the case of a quantum scalar field $\Phi(x)$
linearly interacting with   a classical time dependent  source
$j(x)$ \cite{kar,Bus}. This choice permits to localize the source
within an arbitrary region of spacetime avoiding  the problems of
locality connected to the use of bare or dressed state in its
quantum description \cite{Mate}.
 The Hamiltonian describing the
system is then:
\begin{equation}
H_{tot}=H_{0}+H_{int}
\end{equation}
with
\begin{eqnarray}\label{kaleo}
H_{0} & = & \frac{1}{2}  \int \frac{\rmd^{3}\textbf{k}}{2\omega}
\,\omega
\left(a^{\dag}(\textbf{k})a(\textbf{k})+a(\textbf{k})a^{\dag}(\textbf{k})\right)\nonumber
\\
H_{int} & = & g\int
_{-\infty}^{+\infty}\rmd^{3}\textbf{x}\Phi(x)j(x)
\end{eqnarray}
where $g$ is the source-field coupling constant.
In this model the time
evolution of the source is assigned and the quantum aspects of the system are  described by
the field quantum state. The equation of motion, in the
interaction picture, is
\begin{equation}\label{evovu}
i\frac{\partial}{\partial t}|t\rangle=H_{int}|t\rangle .
\end{equation}
By taking as initial condition at $t=0$  the source off
$\big(j(\textbf{x},0)=0\big)$ and the field in its vacuum state
$|0\rangle$, the formal solution of the equation~(\ref{evovu}) is:
\begin{equation} \label{opevu}
|t\rangle=U(t)|0\rangle
\end{equation}
where time evolution operator is:
\begin{equation}\label{fersic}
U(t)=T \exp\left(-i\int_{0}^{t}\rmd t' H_{int}(t')\right)
\end{equation}
with   $T $    the time ordering operator.  In our model the
commutator of the interaction  of Hamiltonian at two different
times is a c-number given by:
\begin{equation}
 \fl [H_{int}(t_{1}),H_{int}(t_{2})]=ig^{2}\int
\rmd^{3}\textbf{x}_{1}\int
\rmd^{3}\textbf{x}_{2}\Delta(\textbf{x}_{1}-\textbf{x}_{2},t_{1}-t_{2})j(\textbf{x}_{1},t_{1})j(\textbf{x}_{2},t_{2})
\end{equation}
where the propagator  $\Delta$ function is given in terms of the
field commutator as \cite{Bjo2}:
\begin{equation}[\Phi(x),\Phi(y)]=i\Delta(x-y).
\end{equation}
$\Delta(x)$  is real and is zero when its argument is spacelike
($x^{2}<0$). This allows \cite{Zur} to write the $U(t)$ as
\begin{equation}\label{fersic2}
U(t)=\exp\left(-i\int_{0}^{t}\rmd t' H_{int}(t')\right)e^{-\xi(t)}
\end{equation}
where
\begin{eqnarray} \label{stefrub}
\lefteqn{\xi(t)=\frac{1}{2}\int_{0}^{t}\rmd t_{1}\int_{0}^{t}\rmd t_{2}[H_{int}(t_{1}),H_{int}(t_{2})]\Theta(t_{1}-t_{2}){}}\nonumber\\
{} & &=\frac{ig^{2}}{2}\int_{0}^{t}\!\!\rmd
t_{1}\int_{0}^{t}\!\!\rmd t_{2}{}\nonumber
\\
{}& & \times\int\!\!\rmd^{3}\textbf{x}_{1}\int\!\!
\rmd^{3}\textbf{x}_{2}\Delta(\textbf{x}_{1}-\textbf{x}_{2},t_{1}-t_{2})j(\textbf{x}_{1},t_{1})
j(\textbf{x}_{2},t_{2})\Theta(t_{1}-t_{2}).
\end{eqnarray}
Because $\Delta(x)$ and $j(x)$ are real, $\xi(t)$ given by the
expression~(\ref{fersic2}) is an ordinary imaginary function.

 By using the decomposition of the field $\Phi(x)$ in terms of its positive
  and negative frequency parts, as given in equation~(\ref{modena}), it  is possible to transform
   the evolution operator  $U(t)$ in the form \cite{Zur}:
\begin{equation}\label{fersic3}
\fl U(t)=\exp\left(-i\int_{0}^{t}\rmd t'
H_{int}^{-}(t')\right)\exp\left(-i\int_{0}^{t}\rmd t'
H_{int}^{+}(t')\right)e^{-\xi(t)}e^{\alpha(t)}
\end{equation}
where
\begin{equation}\label{milto2}
H_{int}^{\pm}(t)= g\int
\rmd^{3}\textbf{x}\Phi_{\pm}(\textbf{x},t)j(\textbf{x},t)
\end{equation}
and
\begin{eqnarray}\label{passan}
 \fl \alpha(t)&=&-\frac{1}{2}\Bigg[-i\int_{0}^{t}\rmd t'
H_{int}^{-}(t'),-i\int_{0}^{t}\rmd t''
H_{int}^{+}(t'')\Bigg] \nonumber \\
\fl  &=&\frac{ig^{2}}{2}\int_{0}^{t}\rmd t_{1}\int_{0}^{t}\rmd
t_{2}\int \rmd^{3}\textbf{x}_{1} \int
   \rmd^{3}\textbf{x}_{2}j(\textbf{x}_{1},t_{1})\Delta_{-}(\textbf{x}_{1}-\textbf{x}_{2},t_{1}-t_{2})j(\textbf{x}_{2},t_{2})
\end{eqnarray}
where  $\Delta_{-}$ is the negative frequency  part of the
propagator  $\Delta$ given by
\begin{equation}[\Phi_{-}(x),\Phi_{+}(y)]=i\Delta_{-}(x-y).
\end{equation}
 The exponent  $\alpha(t)$ in the expression~(\ref{fersic3}) is an ordinary
 function and  the equation~(\ref{passan}) gives  its dependence from the
 source $j(x)$.
 The form~(\ref{passan}) of the evolution operator makes possible
 to obtain simply the time evolution from any initial state. In fact with  our initial condition the
 equations~(\ref{opevu}) and  (\ref{fersic3}) give the state of the system as:
\begin{equation}\label{vizzo1}
|t\rangle=U(t)|0\rangle=e^{\alpha(t)}e^{-\xi(t)}\exp\left(-i\int_{0}^{t}\rmd
t' H_{int}^{-}(t')\right)|0\rangle
\end{equation}
where use has been made of the fact that $H_{int}^{+}(t')$
contains only annihilation operators.
The state   $|t\rangle$  is not a single particle state, however its single particle component can be shown
to behave non locally as was the case of
the single particle state generated previously by the application of the free field operator on the vacuum.
In fact projecting the state  $|t\rangle$ on the one-particle quantum state
$| \textbf{x}\rangle= a^{\dag}(\textbf{x})| 0 \rangle$   we get the single particle amplitude transition
 $\psi(x)$:
\begin{eqnarray}\label{acara}
\psi(x)=\langle 0|a(\textbf{x})|t\rangle &=&g\int_{0}^{t} \rmd t'\int
\rmd^{3}\textbf{x}'j(\textbf{x}',t')\Delta_{+}(\textbf{x}-\textbf{x}',t-t')\nonumber \\
 & =& g\int_{0}^{t} \rmd t'\int
\rmd^{3}\textbf{x}'j(\textbf{x}',t') \langle 0 | \Phi(x)\Phi(x')|
0 \rangle .
\end{eqnarray}
From the properties of $\Delta_{+}$  it is immediate to deduce
from the expression~(\ref{acara}) that the  single particle
amplitude probability in the case of the state generated from the
source develops non zero contribution outside the light cone. This
behaviour is connected to the corresponding one of the free field
transition amplitude given by the equation~(\ref{berluska}).
 This result is again in agreement with  the Hegerfeldt's
 theorem and with previous results on the non local behaviour of relativistic wavefunctions \cite{Bus,Bar}.

Moreover we note that  expression   in equation~(\ref{acara}) for
the single particle amplitude probability is exact to all order.
If the time spacetime region where the source is different from
zero is an infinitesimal region  around the spacetime point
$y=(\textbf{y},y_{0})$, we have  $j(x)=\delta^{4}(x-y)$ and
substituting  in equation~(\ref{acara}) we obtain:
\begin{equation}
\psi(x)=g\,\Theta(t-y_{0})\Delta_{+}(\textbf{x}-\textbf{y}',t-y_{0}).
\end{equation}
This result is identical to the one obtained previously for
pointlike sources within the first  order of perturbation
theory \cite{Bus}.

\section{Expectation values of one-point ``good operators''}\label{intro3}
As shown before non local behaviour of single particle amplitudes
does  not appear to be  a good test of non locality in QFT being
built from scalar product of spatially extended states or from
states whose projection do not satisfy the microcausality
principle. In view of the previous considerations  it remains the
question if non local  behaviours  can be observed. To give a
meaning  to this question it is necessary to examine  not only the
generation but also the other side of the complete process that is
the detection.
 It has been previously shown \cite{Bus},  within second order perturbation theory, that the calculation of  averages,
 on states generated by pointlike sources,
of one-point localization operators  give results that cannot be
interpreted as the presence of non local properties if these
operators do not satisfy the microcausality principle  \cite{Bus}.
Therefore in the following we shall consider the average values of
``good'' one-point operators  satisfying  the microcausality
principle and  that are  expressed in terms of the field operator
and its space and time derivatives at a given spacetime point. The
result  will be exact and  valid for any ``good'' operator.

\begin{figure}[tpb]
  \begin{centering}
    \includegraphics[width=\textwidth]{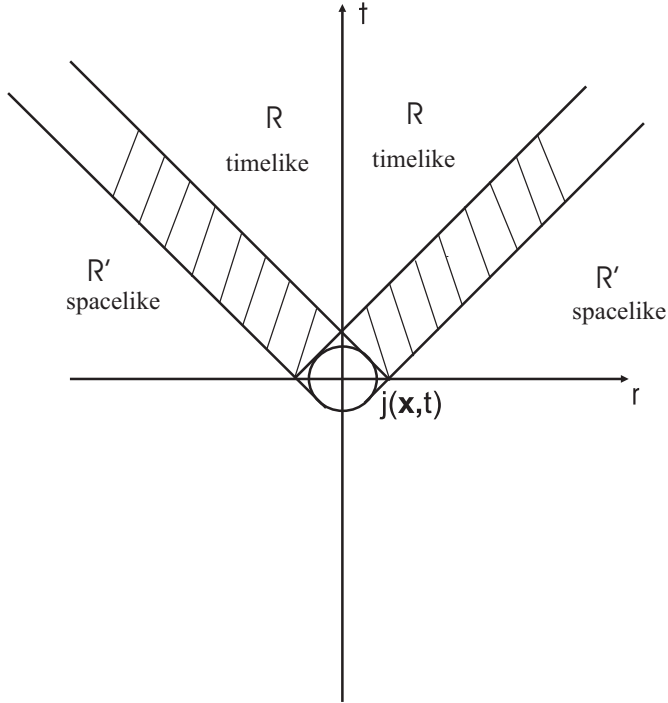}
   \caption{\label{fig:con3}Source forward lightcone: the effects present in separation regions between R and R' may be
   local for a part of source and  non local for another.}
 \end{centering}
\end{figure}

 The adjoint of the time evolution operator $U(t)$
  is  immediately obtained as:
\begin{equation}\label{foix}
\fl
U^{\dag}(t)=U^{-1}(t)=e^{\xi(t)}e^{-\alpha(t)}\exp\left(i\int_{0}^{t}\rmd
t' H_{int}^{+}(t')\right)\exp\left(i\int_{0}^{t}\rmd t'
H_{int}^{-}(t')\right).
\end{equation}  At  first we shall  consider the expectation value of the field operator
   $\Phi(x)$  on the state generated by the source  $|t\rangle$,
   which is given by:
\begin{eqnarray}\label{meidocampo2}
\langle t |\Phi(x)|t\rangle&=&\langle 0
|U^{-1}(t)\Phi(x)U(t)|0\rangle\nonumber \\
&=&\langle 0
|U^{-1}(t)\left(U(t)\Phi(x)+\left[\Phi(x),U(t)\right]\right)|0\rangle
.
\end{eqnarray}
The commutator appearing  above    can be transformed, using
equations~(\ref{kaleo}) and (\ref{fersic3}), as
\begin{eqnarray}\label{fiorva2}
[\Phi(x),U(t)] & = & -i U(t)\int_{0}^{t}
\rmd t'[\Phi(x),H_{int}(t')]\nonumber \\
& = & U(t)g\int_{0}^{t} \rmd t'\int
\rmd^{3}\textbf{x}'j(\textbf{x}',t')\Delta(\textbf{x}-\textbf{x}',t-t')
\end{eqnarray}
 Thus, taking into account that $\langle 0
|\Phi(x)|0\rangle=0$ and introducing the
retarded propagator $\Delta^{ret}(x)=\Theta(t)\Delta(x)$, the expectation value (\ref{meidocampo2})
takes the form:
\begin{equation}\label{meidocampo}
\langle t |\Phi(x)|t\rangle  =  g\int_{0}^{t} \rmd t'\int
\rmd^{3}\textbf{x}'j(\textbf{x}',t')\Delta_{ret}(\textbf{x}-\textbf{x}',t-t')=
g \widetilde{\Delta}_{R}(x-y).
\end{equation}
In  equation~(\ref{meidocampo}) it is $t>t'>0$, while
$(x-y)\equiv(\textbf{x}-\textbf{y},t-y_{0})$ and  ($\textbf{y}$,
$y_{0}$) indicates a spacetime point inside the spacetime region
where $j(\textbf{x},t)\neq 0$.  Therefore the function
$\widetilde{\Delta}_{R}(x-y)$  defined by the
expression~(\ref{meidocampo}) contains the  spacetime evolution of
the source and, because of its construction,  is zero outside a
forward lightcone containing the spacetime region where the source
is non zero. This lightcone has a vertex in the part of the
spacetime region where $j(\textbf{x},t) = 0$
(Figure~\ref{fig:con3}).

As a consequence of  the linearity of the
equation of motion of the quantum scalar field,  the expression
for  $\langle \Phi(x)\rangle$ is identical to the retarded
solution obtained by a classical scalar field satisfying the
equation of motion
\begin{equation}\left(
\frac{\partial^{2}}{\partial
t^{2}}-\nabla^{2}+m^{2}\right)\Phi(\textbf{x},t)=g\,j(\textbf{x},t)
\end{equation}
where  $j(\textbf{x},t)$ is the same  source
appearing in the quantum equations. For  a classical scalar source
differing  from zero only in an infinitesimal spacetime region
around  $y=(\textbf{y},y_{0})$, it is
$j(x)=\delta^{4}(x-y)$. In this case the equation of motion
(\ref{meidocampo}) gives for the average scalar field
\begin{equation}
\langle  \Phi(x) \rangle=g \Delta_{ret}(x-y).
\end{equation}
This result is exact and  corresponds to  the one obtained
previously within the second
order of perturbation theory   in the case of  pointlike sources \cite{Bus}. This shows that, at
least for the field expectation value, the result obtained  via
second order perturbation theory is  exact.

We shall
consider the expectation value of a generic  operator $\widehat{O}(x)$
that may be expressed  as an arbitrary analytical function
 of the field operator and of its space and time derivatives. Its expansion  in
power series being
\begin{equation}\label{martins}
\widehat{O}(\Phi(x),\dot{\Phi}(x),\nabla\Phi(x))=\sum_{j,k,l}c_{ j
k l}\Phi^{j}(x)\dot{\Phi}^{k}(x)(\nabla\Phi)^{l}(x)
\end{equation}
with $c_{ j,k l}$ ordinary coefficients.
To obtain  the average value of $\widehat{O}(x)$ we have to evaluate
the expectations value of the powers and products of the field operator and its
time and space derivatives.

First we shall consider   the expectation value, on the state
$|t\rangle$ given by  equation~(\ref{opevu}), of the m-power of
the field operator  $\Phi^{m}(x)$ . Using the explicit calculation
of \ref{sec:cel1} we find:
\begin{eqnarray}\label{corleo}
\langle t |\Phi^{m}(x)|t\rangle & = &  \langle 0 |(g
\widetilde{\Delta}_{R}(x-y)+\Phi(x))^{m}|0\rangle\nonumber \\
& = & \sum_{\nu=0}^{m}{m \choose
\nu}\left(g\widetilde{\Delta}_{R}(x-y)\right)^{\nu}\langle 0|
\Phi^{m-\nu}(x)|0\rangle
\end{eqnarray}
 In  the expression~(\ref{corleo}) the first term (for $\nu=0$) $\langle 0
|\Phi^{m}(x)|0\rangle$ is the only  one independent   from the
field source coupling constant $g$. Moreover it differs from zero
over the whole  spacetime and it represents the expectation values
of $\Phi^{m}(x)$ on the vacuum state. The remaining terms are
source dependent  and, because  of the presence  of the term  $g
\widetilde{\Delta}_{R}(x-y)$, causally retarded. In order to
examine the causal effects due to the variation of the source the
vacuum contribution must therefore be subtracted in the
expression~(\ref{corleo}) in agreement  with previous results
\cite{Com3,Com2,Com1,Ber2}.

Equation~(\ref{corleo}) has a simple interpretation: the
expectation value of the m-th power of the field operator in our
system is equal to  the m-th power of the field operator on the
vacuum shifted by $g\widetilde{\Delta}_{R}(x-y)$. This result of
the expression~(\ref{corleo}) has been obtained independently from
the explicit expression of the initial state of the field and
utilizing  only relationship between $\Phi^{m}(x)$, the time
evolution operator $U(t)$ and its adjoint $U^{\dag}(t)$. Thus the
result given by the equation~(\ref{corleo}) can be extended to:
\begin{equation}
\langle t_{i} |\Phi^{m}(x)|t_{i}\rangle=\langle i |g
\widetilde{\Delta}_{R}(x-y)+\Phi(x))^{m}|i\rangle
\end{equation}
where now $|i\rangle $  is an arbitrary initial state for the field
and  $|t_{i}\rangle=U(t)|i\rangle$.

We shall now proceed to calculate the expectation value on our
state $|t\rangle$ of the  gradient of the field operator
$\nabla_{\textbf{\scriptsize{x}}}\Phi(x)$. It is:
\begin{eqnarray}\label{bulbar}
\langle t
|\nabla_{\textbf{\scriptsize{x}}}\Phi(x)|t\rangle&=&\langle 0
|U^{-1}(t)\nabla_{\textbf{\scriptsize{x}}}\Phi(x)U(t)|0\rangle\nonumber \\
&=&\langle 0
|U^{-1}(t)\left(U(t)\nabla_{\textbf{\scriptsize{x}}}\Phi(x)+
\left[\nabla_{\textbf{\scriptsize{x}}}\Phi(x),U(t)\right]\right)|0\rangle
.
\end{eqnarray}
The commutator appearing above   can easily obtained  by noting
that  $\nabla_{\textbf{\scriptsize{x}}}$ acts only on space
coordinate. Thus  using also the equations~(\ref{fiorva2}) and
(\ref{meidocampo}), we get:
\begin{equation}\label{voce}
\left[\nabla_{\textbf{\scriptsize{x}}}\Phi(x),U(t)\right]=\nabla_{\textbf{\scriptsize{x}}}\left[\Phi(x),U(t)\right]=
g\,U(t)\nabla_{\textbf{\scriptsize{x}}}
\widetilde{\Delta}_{R}(x-y)
\end{equation}
and by substituting this in  the expression~(\ref{bulbar}), we
obtain at the end:
\begin{equation}
\langle t|
\nabla_{\textbf{\scriptsize{x}}}\Phi(x)|t\rangle=g\nabla_{\textbf{\scriptsize{x}}}
\widetilde{\Delta}_{R}(x-y)
\end{equation}
where use has been made of $\langle 0|
\nabla_{\textbf{\scriptsize{x}}}\Phi(x)|0\rangle=0$. The
expectation value of $\nabla_{\textbf{\scriptsize{x}}}\Phi(x)$ is
thus simply expressed in terms of the gradient
$\widetilde{\Delta}_{R}(x-y)$; it is therefore again retarded and
causal. The expectation value of the m-th power of
$\nabla_{\textbf{\scriptsize{x}}}\Phi(x)$ may be obtained
following  with the appropriate changes  the procedure followed to
calculate  the expectation value of $\Phi^{m}(x)$. We get in this
case:
\begin{eqnarray}\label{3b}
\fl \langle t
|(\nabla_{\textbf{\scriptsize{x}}}\Phi(x))^{m}|t\rangle & = &
\langle 0
|\left(\nabla_{\textbf{\scriptsize{x}}}\Phi(x)+g\nabla_{\textbf{\scriptsize{x}}}
\widetilde{\Delta}_{R}(x-y)\right)^{m}|0\rangle\nonumber \\ \fl &
= & \sum_{\nu=0}^{m}{m \choose
\nu}\left(g\nabla_{\textbf{\scriptsize{x}}}
\widetilde{\Delta}_{R}(x-y)\right)^{\nu}\langle 0|
(\nabla_{\textbf{\scriptsize{x}}}\Phi(x))^{m-\nu}|0\rangle .
\end{eqnarray}

As last we proceed to calculate the expectation value of time
derivative of the field operator $\partial_{t}\Phi(x)$ and of its
m-th power. The expectation value of $\partial_{t}\Phi(x)$ on the
state $|t\rangle$ is
\begin{eqnarray}\label{lenny}
\langle t|\partial_{t}\Phi(x)|t \rangle&=&\langle
0|U^{-1}\partial_{t}\Phi(x)U(t)|0 \rangle\nonumber
\\&=&\langle 0
|U^{-1}(t)\left(U(t)\partial_{t}\Phi(x)+\left[\partial_{t}\Phi(x),U(t)\right]\right)|0\rangle
.
\end{eqnarray}
 In this case  to calculate  the commutator appearing in
 expression~(\ref{lenny}) use can be made of  the equations~(\ref{kaleo}) and
(\ref{fersic3}) obtaining:
\begin{eqnarray}\label{lenny2}
\fl [\partial_{t}\Phi(x),U(t)] & = & -i U(t)\int_{0}^{t}
\rmd t'[\partial_{t}\Phi(x),H_{int}(t')]\nonumber \\
\fl & = & -i U(t)\Big\{\frac{\partial}{\partial
t}\Big(\int_{0}^{t} \rmd t'[\Phi(x),H_{int}(t')]\Big)
-[\Phi(x),H_{int}(t)]\Big\}
\nonumber \\
\fl & = &U(t) g\partial_{t}
\widetilde{\Delta}_{R}(\textbf{x}-\textbf{y},t-y_{0})
\end{eqnarray}
where  it has been exploited the property that $\Delta(x)$ is zero when its argument is spacelike.
 Substituting the expression~(\ref{lenny2}) in (\ref{lenny}), we get
\begin{equation}\label{lenny3}
\langle t|\partial_{t}\Phi(x)|t\rangle=g\partial_{t}
\widetilde{\Delta}_{R}(x-y)
\end{equation}
where  $\langle 0|\partial_{t}\Phi(x)|0\rangle=0$ has been taken
into account. The expectation value of $\partial_{t}\Phi(x)$ in
 equation~(\ref{lenny3}) is expressed as time derivative of
$\widetilde{\Delta}_{R}(x-y)$ and is thus causally retarded.
Similarly to the previous results the expectation value on
$|t\rangle$ of $(\partial_{t}\Phi)^{m}(x)$ is:
\begin{eqnarray} \label{rap3}
\langle t |(\partial_{t}\Phi)^{m}(x)|t\rangle &=&\langle 0
|(\partial_{t}\Phi(x)+g\partial_{t}
\widetilde{\Delta}_{R}(x-y)^{m}|0\rangle\nonumber \\
& = & \sum_{\nu=0}^{m}{m \choose \nu}\left(g\partial_{t}
\widetilde{\Delta}_{R}(x-y)\right)^{\nu}\langle 0|
(\partial_{t}\Phi^{m-\nu}(x)|0\rangle
\end{eqnarray}
The results obtained in equations~(\ref{corleo}), (\ref{3b}) and
(\ref{rap3})   can be used to evaluate  the expectation value of
the  operator $\widehat{O}(\Phi(x),\dot{\Phi}(x),\nabla\Phi(x))$
analytical  function of field operator and its derivatives. In
particular the average value of a generic term of the power series
of equation~(\ref{martins}), takes the form
\begin{eqnarray}\label{messin}
\fl \langle t
|\Phi^{j}(x)\dot{\Phi}^{k}(x)(\nabla\Phi)^{l}(x)|t\rangle=
\nonumber\\ \fl  \langle 0 |U^{-1}(t)\Phi^{j}(x)U (t)U^{-1}(t)
\dot{\Phi}^{k}(x)U(t)
U^{-1}(t)(\nabla\Phi)^{l}(x)U(t)|0\rangle= \nonumber\\
\fl  \langle 0|\Big (\Phi(x)+g
\widetilde{\Delta}_{R}(x-y)\Big)^{j}\Big
(\dot{\Phi}(x)+g\partial_{t} \widetilde{\Delta}_{R}(x-y)\Big)^{k}
\Big (\nabla\Phi(x)+g\nabla
\widetilde{\Delta}_{R}(x-y)\Big)^{l}|0\rangle
\end{eqnarray}
 In this case we  observe that the
expectation value, on the field vacuum, of the  term appearing in
(\ref{messin}), $\langle 0
|\Phi^{j}(x)\dot{\Phi}^{k}(x)(\nabla\Phi)^{l}(x)|0\rangle$, is
source independent  and in general not zero all over the
spacetime. All the remaining terms contain  at least a factor
$g\widetilde{\Delta}_{R}(x-y)$ and  its space or time derivative;
therefore they depend on the source and are causally retarded.
Thus the expectation value of any observable
$\widehat{O}(\Phi(x),\dot{\Phi}(x),\nabla\Phi(x))$,  that may be
expanded as in the expression (\ref{martins}) shows  a causal
evolution if its expectation value on the field vacuum is
subtracted. This result
 is  valid to all orders of the coupling constant and  is the
main result of this paragraph. It extends for our QFT model the previous
results \cite{kar,Com3,Ber2,Com2,Com1}  obtained at finite
order of perturbation theory.

Using equation~(\ref{messin}) the expectation value of the
observable $\widehat{O}(\Phi(x),\dot{\Phi}(x),\nabla\Phi(x))$ on
$|t\rangle$ may be explicitly written  as:
\begin{eqnarray}
\fl \quad \langle
t|\widehat{O}(\!\Phi(x),\dot{\Phi}(x),\nabla\Phi)(x)|t\rangle=  \nonumber \\
 \fl \qquad \langle
 0 |\widehat{O}\!\left(\!\Phi(x)+ g \widetilde{\Delta}_{R}(x-y),\dot{\Phi}(x)+g \partial_{t}\widetilde{\Delta}_{R}(x-y),\nabla\Phi(x)+g\nabla \widetilde{\Delta}_{R}(x-y)\right)\!\!|0\rangle
\end{eqnarray}
The above expression is one of the main results of our paper and shows that  the   expectation value of  $\widehat{O}$ on
$|t\rangle$ coincides with the expectation value on the field
vacuum of the same expression in terms of operator $\Phi(x)$ and
its derivative shifted by $g \widetilde{\Delta}_{R}(x-y)$ and its
corresponding derivatives.

The above results permit to obtain  immediately the explicit form
of the expectation value of any  arbitrary one-point observable
represented as a function of the field operator and its space and
time derivatives. In the following we shall apply this result to
the energy density of the field defined by the
equation~(\ref{dink}). Following   the above recipes its
expectation value on $|t\rangle$  is  immediately given by:
\begin{eqnarray}\label{bertallofe}
\fl \langle t |\mathcal{H}(x)|t\rangle &=&\langle 0
|\mathcal{H}(x)|0\rangle\nonumber\\
\fl
&+&1/2\Bigg(\Big(\partial_{t}{g\widetilde{\Delta}}_{R}(x-y)\Big)^{2}+m^{2}\Big(g\widetilde{\Delta}_{R}(x-y)\Big)^{2}+
\Big|\nabla _{x}g\widetilde{\Delta}_{R}(x-y)\Big|^{2}\Bigg)
\end{eqnarray}
where the expectation values  on  the vacuum of the field and its
derivatives are zero. The first term of
expression~(\ref{bertallofe}) represents the vacuum contribution
to energy density, the second, as expected from the previous
discussion, is source dependent and causally retarded. Although
the two point correlation function on $ |t \rangle  $ is not of
the form~(\ref{messin}) depending  on the field at two separate
spacetime points, it is also possible to obtain  its expectation
value on $|t\rangle $ as:
\begin{eqnarray}\label{cammar}
 \fl {}\langle t |\Phi (x)\Phi (x')|t\rangle & = &\langle 0
|U^{-1}(t)\Phi(x)U(t)U^{-1}(t)\Phi(x')U(t)|0\rangle  \nonumber \\
 & =&\langle 0 |\Big(\Phi
(x)+g\widetilde{\Delta}_{R}(x-y)\Big)\Big(\Phi(x')+g\widetilde{\Delta}_{R}(x'-y)\Big)|0\rangle\nonumber
\\  &= &\langle 0 |\Phi
(x)\Phi(x')|0\rangle+g^{2}\widetilde{\Delta}_{R}(x-y)\widetilde{\Delta}_{R}(x'-y).
\end{eqnarray}
It appears  from  the expression~(\ref{cammar}) that it again
consists of two parts.  The first, which  is the vacuum zero point
correlation function, is not zero at spacelike distances $x-x'$
being  a well known peculiarity of the field vacuum state
\cite{Bjo2}, the second is source dependent and also not zero for
some ranges of  $x-x'$ spacelike. The physical explanation of this
behaviour is that in these regions the field at $x$ and $x'$ are
both causally correlated with the source and this induces a
correlation for field at these points so this is in agreement with
causality requirements.

\section{Conclusions}\label{intro4}

The problem of causal propagation and  the appearance of non locality have widely been investigated both non relativistic QED, 
and in relativistic quantum Mechanics, where however the concept of localization appears questionable, even if most analyses have been limited to the lowest order of perturbation theory.  
Non locality is present in QFT even for free fields case  and the question is raised if it may  lead to the violation of Einstein's causality.  
A satisfactory study of causality, which must be exact, and the appearance of non local effects would require however the exact solution of the model \cite{Pow}
 under examination. It appears thus of interest to study the causal behaviour with models of matter-field interaction that permit  to evaluated the time evolution of the quantities of interest at all orders.

 In the first part of paper we have investigated  nonlocality   present in the free QFT in the case of a scalar quantum field for two-point functions and single particle wave amplitude. 
The states are obtained either by the action of the field operator on the vacuum or through the action of the Newton Wigner localization operator.
 This analysis has  lead to the conclusion that it is inappropriate consider the states so generated as being effectively point like  localized in a point in space even if they may be 
characterized by a position parameter. In fact states with different position parameters present a spatial overlap and the expectation values on them of a ``good''  local observable, 
as field density energy,  can be seen to extend over the whole space. 
The appearance of states with spatially extended characteristics  is responsible of the non locality present in the two-point function that gives the Feynman propagator. 
The development of non locality during the time evolution of field states, even if initially localized even in the Newton Wigner sense,  can be viewed as a consequence of the Hegerfeldt's theorem. 
This behaviour, because of its connection with spatially extended states,  may   not be interpreted as failure of causal propagation in QFT \cite{kar,Rubin}.

The  appearance of non locality for free fields requires to establish if it is observable and thus to lead to examination of models of matter-fields interaction 
where fields may be generated and absorbed. In this paper we have in particular considered a simple QFT model consisting of a quantum scalar field linearly interacting with
a classical source localized in a finite space-time region. The choice of a classical
source permits also to avoid the appearance of spurious non local effects linked to the
difficulty to localize a quantum mechanical source. Moreover the simplicity of the system has allowed to calculate exactly   time evolution of the state of the system and 
the expectation values of any ``good'' local observable, represented by single-point operators analytical function of the field and its space and time derivatives. 
We have found that non locality is present in the expectation value of the both single-point local operators and of two-point correlation function.
Moreover the non local terms appearing in the expectation values are source independent and can be seen to be connected to zero point vacuum fluctuations 
where are  already present non local features, the source dependent parts propagate instead in a causal way. 
Thus in order to analyze the causal contributions linked to the variations of the sources, it must be required  
that the contributions arising from the vacuum should be subtracted  in agreement with what suggested in previous works \cite{Ber2,Bus,Com3,Com2}. 
Our results, for any ``good'' local observable, are valid at all orders thus they could  be taken as representative of the causal behaviour of  
interacting QFT model and can be in particular considered as the guidelines to investigate the causal propagation in more complex models of matter-scalar field interaction.

At the end even if in our system the expectation value of  any ``good'' local quantity appears to depend causally  from the source, 
the presence of non local parts in it gives rise to questions about the  possibility of their measurement and detection.  
In order to give a concrete meaning of the observability of non locality in matter-field interactions one should analyze suitable quantum detector models  \cite{Gla1,UNR}. 
The response of this kind of detectors in the field detection should permit to investigate better the relation between the causal propagation 
and locality and also discuss more physically the applications of Hegerfeldt's theorem to localized field states.

\ack We wish to dedicate this paper to the memory of A.E. Power,
who first suggested to one of the authors (G.C.) the importance to
consider exactly solvable models in the analyses of causality and
non locality in QFT.

\appendix
\section{Expectation value of $\Phi^{m}(x)$}\label{sec:cel1}
Here we shall give the explicit calculation of the field m-th
power in state $|t\rangle$ given by equation~(\ref{opevu}) and
that represents the exact solution of the scalar field-source
linearly coupling. The expectation value may be written as:
\begin{eqnarray}\label{fiq99} \langle t
|\Phi^{m}(x)|t\rangle&=&\langle 0
|U^{-1}(t)\Phi^{m}(x)U(t)|0\rangle\nonumber \\
&=&\langle 0
|\left(\Phi^{m}(x)U^{-1}(t)+[U^{-1}(t),\Phi^{m}(x)]\right)U(t)|0\rangle
\end{eqnarray}
 where the operators $U(t)$ and  $U^{-1}(t)$
have the form given by equations~(\ref{fersic3}) and (\ref{foix}).
To evaluate  the commutator appearing in the
expression~(\ref{fiq99}), we show for induction that:
\begin{equation}\label{commut}
[U^{-1}(t),\Phi^{m}(x)]=\sum_{k=1}^{m}{m \choose
k}\left(g\widetilde{\Delta}_{R}(x-y)\right)^{k}\Phi^{m-k}(x)U^{-1}(t)
\end{equation}
In fact for $m=1$ from (\ref{commut}) we get
\begin{equation}\label{commuta}
[U^{-1}(t),\Phi(x)]=g\widetilde{\Delta}_{R}(x-y)U^{-1}(t)
\end{equation}
that is the expression one get by explicitely calculating the
commutator using  equations (\ref{kaleo})  and
(\ref{foix}).\\Assuming that  the expression~(\ref{commut}) is
true for a generic $n$ the term $n+1$ may be obtained as:
\begin{eqnarray}
\fl[U^{-1}(t),\Phi^{n+1}(x)]=[U^{-1}(t),\Phi^{n}(x)]\Phi(x)+\Phi^{n}(x)[U^{-1}(t),\Phi(x)]
\nonumber \\
=\sum_{k=1}^{n}{n \choose
k}\left(g\widetilde{\Delta}_{R}(x-y)\right)^{k}\Phi^{n-k}(x)U^{-1}(t)\Phi(x)+\Phi^{n}(x)g\widetilde{\Delta}_{R}(x-y)
U^{-1}(t){}\nonumber\\  =\sum_{k=1}^{n}{n \choose
k}\left(g\widetilde{\Delta}_{R}(x-y)\right)^{k}\Phi^{n-k+1}(x)U^{-1}(t)\nonumber\\
 +\sum_{k=1}^{n}{n \choose
k}\left(g\widetilde{\Delta}_{R}(x-y)\right)^{k+1}\Phi^{n-k}(x)U^{-1}(t){}+\Phi^{n}(x)g\widetilde{\Delta}_{R}(x-y)
U^{-1}(t)\nonumber\\
\end{eqnarray}
The term proportional to $\Phi^{n}$ can be written as
\begin{eqnarray}
\lefteqn{{n \choose
1}g\widetilde{\Delta}_{R}(x-y)\Phi^{n}(x)U^{-1}(t)+g\widetilde{\Delta}_{R}(x-y)\Phi^{n}(x)U^{-1}(t)=}{}\nonumber
\\ & & {}={n+1 \choose 1}g\widetilde{\Delta}_{R}(x-y)\Phi^{n}(x)U^{-1}(t)
\end{eqnarray}
while the one proportional to  $\Phi^{n+1-\nu}$ as
\begin{eqnarray}
 \lefteqn{{n\choose \nu}
\left(g\widetilde{\Delta}_{R}(x-y)\right)^{\nu}\Phi^{n+1-\nu}(x)U^{-1}(t)}{}\nonumber
\\  & &{} +{n \choose
\nu-1}\left(g\widetilde{\Delta}_{R}(x-y)\right)^{\nu}\Phi^{n+1-\nu}(x)U^{-1}(t)\nonumber
\\  &  &={n+1 \choose
\nu}\left(g\widetilde{\Delta}_{R}(x-y)\right)^{\nu}\Phi^{n+1-\nu}(x)U^{-1}(t)
\end{eqnarray}
At last the term where it appears $\Phi^{0}=1$ as
\begin{eqnarray}
\lefteqn{{n\choose
n}g\widetilde{\Delta}_{R}^{n+1}(x-y)U^{-1}(t)=g\widetilde{\Delta}_{R}^{n+1}(x-y)U^{-1}(t)=}{}\nonumber
\\ & & {}={n+1\choose n+1}g\widetilde{\Delta}_{R}^{n+1}(x-y)U^{-1}(t)
\end{eqnarray}
 Thus equation~(\ref{commut}) is true for $n+1$. Thus we may write:
\begin{eqnarray}\label{mikkiuo}
\lefteqn{U^{-1}(t)\Phi^{m}(x)U(t)=\bigg(\Phi^{m}(x)U^{-1}(t)+[U^{-1}(t),\Phi^{m}(x)]\bigg)U(t)={}}
\nonumber
\\ {} & &\quad\Phi^{m}(x)+\sum_{\nu=1}^{m}{m \choose
\nu}\left(g\widetilde{\Delta}_{R}(x-y)\right)^{\nu}\Phi^{m-\nu}(x)=\nonumber\\
 {} & &\quad(g\widetilde{\Delta}_{R}(x-y)+\Phi(x))^{m}
\end{eqnarray}
where the Newton formula for
$g\widetilde{\Delta}_{R}(x-y)+\Phi(x))^{m}$ has been used.
Inserting the expression~(\ref{mikkiuo}) in (\ref{fiq99}), we
obtain:
\begin{equation}
\langle t |\Phi^{m}(x)|t\rangle=\langle 0
|(g\widetilde{\Delta}_{R}(x-y)+\Phi(x))^{m}|0\rangle
\end{equation}

\end{document}